# Unlocking New Frontiers: The Versatile Potential of the Brazilian Multipurpose Reactor

Luiz P. de Oliveira[1,*], Alexandre P.S. Souza[1,2], Carlos G.S. Santos[1,2], Iberê R.S. Júnior[1], Barbara P.G. Silva[1], Marco A.S. Pereira[1], Frederico A. Genezini[2], and José A. Perrotta[1]

[1]**Reator Multipropósito Brasileiro, RMB/CNEN, Av. Prof. Lineu Prestes, 2242 – Cidade Universitária – CEP 05508-000, São Paulo-SP-Brazil**

[2]**Instituto de Pesquisas Energéticas e Nucleares, IPEN/CNEN, Av. Prof. Lineu Prestes, 2242 – Cidade Universitária – CEP 05508-000, São Paulo-SP-Brazil**

[*]luiz.o-fpatria@ipen.br

## Abstract

The multipurpose nature of nuclear research reactors has been investigated. This class of reactors has gained momentum within the international community, given the multiple benefits of nuclear technology in medicine, agriculture, industry, and the development of new materials, being characterized by the simultaneous use of several applications. In this context, Brazil has joined the group of countries with multipurpose reactor projects. The Brazilian Multipurpose Reactor (RMB) will be Brazil's new neutron source, with a thermal power of 30 MW, with a core flux on the order of $\phi = 4 \times 10^{14}$ n/cm$^2$s and a power density of 312.5 W/cm³, ensuring the production of radioisotopes, material and fuel irradiation testing, and neutron beam applications. In this work, the demand for neutron fluxes in different irradiation targets is investigated by Monte Carlo simulations under a perturbation caused by a highly neutron-absorbing material, the Gadolinium-157 (Gd-157) isotope. The results show that Molybdenum-99 (Mo-99) production is not affected by the Gd-157 target, and the minimum proposed thermal neutron flux for the reactor's activity ($\phi_{min} = 5.1 \times 10^{12}$ n/cm$^2$s) is guaranteed even at maximum operational capacity. The potential production of Terbium (Tb) isotopes for nuclear medicine applications in the RMB is briefly discussed.

# 1 Introduction

A research reactor is a nuclear reactor primarily used for the generation and utilization of neutron flux for research and other purposes. Within the class of research reactors, there are those that encompass multiple simultaneous applications, including material irradiation, radioisotope production, and the use of neutron beams and radiation for scientific and technological research purposes [1]. With the growing demand for radioisotopes for nuclear medicine, the need to test new materials for advanced reactors, and the pursuit of scientific innovation, multipurpose nuclear reactors (MNRs) have become strategic infrastructure for countries that invest in technology, health, energy, and defense.

Given the diverse range of research reactor applications, safety requirements do not need to be applied uniformly across all reactor types. The manner in which compliance with these requirements is demonstrated may vary significantly between a high-power MNR and a very low-power reactor, for which the radiological risk to facility personnel, the public, and the environment is minimal [2]. Over the past four decades, many proposals for the design of new research reactors have been developed, showing significant progress since the first criticality of Chicago Pile-1 in December 1942 [3]. A compilation of multipurpose research reactors currently in operation or under construction/planning, with power levels above 10 MW, is presented in Table 1 (this collection complements the data compiled in [3]). Factors such as the high cost of power reactors, the growing need for material testing under irradiation, the increased demand for medical radioisotopes following the international Mo-99 supply crisis in the late 2000s, and the advanced applications of neutron beams in both hard and soft matter studies have driven the development of new multipurpose research reactor projects on a global scale. An example of this proposal is the Indian 20 MW MNR, aimed at consolidating and expanding the scope of research and development in nuclear and allied sciences [4].

In 2009, Teruel and Uddin proposed a research reactor core design optimized for a 10 MW pool-type that used low-enriched Uranium as the standard fuel, aiming for high thermal neutron fluxes, adequate space in the reflector for experimental facilities, and a satisfactory operational life cycle, using exclusively standard low-enriched Uranium fuel, without compromising other essential features of modern multipurpose facilities [5]. Inspired by the German FRM-II reactor, the proposed core adopted an azimuthally asymmetric cylindrical configuration with inner and outer reflectors.

Table 1 - Multipurpose research reactors in operation or under construction with power levels above/equal 10 MW and built after 1980.

| Country | Name | Power (MW) | Setup type | Reflector | Criticality date |
|---|---|---|---|---|---|
| **In operation** | | | | | |
| Libya | IRT-1 | 10 | Tube | Be | 1981 |
| Indonesia | RSG-GAS | 30 | MTR | Be | 1987 |
| Peru | RP-10 | 10 | MTR | Be, Graphite | 1988 |
| Japan | JRR-3 M | 20 | MTR | $D_2O$ | 1990 |
| Korea | HANARO | 30 | Rod | $D_2O$ | 1991 |
| Algeria | ES-SALAM | 15 | Rod | Graphite, $D_2O$ | 1992 |
| Egypt | ETRR-2 | 22 | MTR | $D_2O$ | 1997 |
| Germany | FRM-II | 20 | Involuted plate | $D_2O$ | 2004 |
| Australia | OPAL | 20 | MTR | $D_2O$ | 2006 |
| China | CARR | 60 | MTR | $D_2O$ | 2010 |
| **Under construction or planned** | | | | | |
| France | RJH | 100 | Curved plate | Be | Construction |
| India | MPRR | 20 | MTR | $D_2O$ | Planned |
| South Africa | MPR | 30 | MTR | Be | Planned |
| Netherlands | PALLAS | 25 | MTR | $D_2O$ | Construction |
| Russian | MBIR | 150 | vibroMOX | - | Construction |
| Argentine | RA-10 | 30 | Plate | $D_2O$ | Construction |
| Brazil | RMB | 30 | MTR | Be/ $D_2O$ | Construction |

A few years later, Seo and Cho introduced a new core design concept for multipurpose research reactors using MTR-type (Material Testing Reactor) fuel with in-core irradiation holes of varying sizes, unlike conventional cores with uniform assembly pitch and hole sizes [3]. Two core models were compared through MCNP (Monte Carlo N Particle) calculations [6]: a compact core featuring a larger irradiation hole for beam tubes, yielding higher thermal neutron flux, and a larger core with both small and large holes optimized for fast and thermal neutron fluxes, respectively. The new cores employ smaller fuel assemblies, enhancing safety by allowing higher coolant flow rates and simplifying the design by using only one type of fuel plate. This innovative concept offers significant advantages for the development of multipurpose research reactors.

A significant portion of the design and development of MNRs in recent decades has been undertaken by the Argentine company INVAP. The company is worldwide recognized for delivering comprehensive turn-key solutions, encompassing all project phases—from conceptual design to assisted reactor operation—while also providing technical training, fuel supply, and regulatory compliance support [7].

Recently, Nguyen *et al.* presented a conceptual design of a 10 MW multipurpose nuclear research reactor (MPRR) loaded with low-enriched Uranium (LEU) VVR-KN fuel type in Vietnam [8]. The 500 kW Dalat Nuclear Research Reactor (DNRR), located in Dalat, Vietnam, is currently the only research reactor in the country. Due to its limited power and experimental facilities, the DNRR cannot meet the growing demands for radioisotope production for medical and industrial applications, nuclear physics experiments, and support for future nuclear power development. For this reason, the country is planning to build a new MPRR.

In the end of 2000's, Brazil and Argentina starts a new research reactors project, both design which are very similar and carried out in collaboration with INVAP, were based on OPAL. The RA-10 is a multipurpose nuclear reactor currently under construction at the Ezeiza Atomic Center in Buenos Aires Province, Argentina [9]. Designed by the National Atomic Energy Commission (CNEA) in collaboration with the company INVAP, this project represents one of the largest state investments in science and technology in Argentina. With a thermal power of 30 MW, the RA-10 will replace the RA-3 reactor, which has been in operation since 1967, and will meet the growing global demand for medical radioisotopes. Additionally, it will provide facilities for materials testing, advanced nuclear fuel research, and neutron beam applications. The reactor is an open pool type, utilizing low-enriched Uranium silicide (MTR) fuel and moderated by light water. Its core is surrounded by a heavy water ($D_2O$) reflector tank, ensuring a high thermal neutron flux. The operational cycle is continuous, lasting 29.5 days, with a weekly Mo-99 production exceeding 2,000 Ci. The project also includes doped Silicon production, with an estimated annual capacity between 30 and 60 tons, addressing approximately 40% of current global demand. Furthermore, the RA-10 will contribute to the development of domestic nuclear fuels and enhance research capabilities in biotechnology, materials science, and drug design. This multipurpose facility is expected to strengthen Argentina's position in the international nuclear research and medical isotope production sectors.

Configured as a multipurpose reactor proposal, RMB is a strategic facility designed to meet Brazil's growing demands in research, radioisotope production, and nuclear technological development [10]. The RMB was conceived as a high-performance multipurpose reactor, utilizing low-enriched Uranium fuel with a design focused on safety, efficiency, and operational versatility. The next section presents the RMB in more detail, as well as the implementation of the reactor in the Monte Carlo simulation environment [6].

**2 The New Brazilian Nuclear Reactor**

RMB will be an open pool type reactor with 30 MW power located in Iperó, a municipality about 100 Km from São Paulo, Brazil. Its civil construction works in 2025 and is expected to reach criticality by 2032 (artistic view of the future facility in Figure 1). The main objective of the project is to make Brazil self-sufficient in radiopharmaceuticals through the production of radioisotopes. In addition, the RMB will also be capable of testing and characterizing nuclear fuels and structural materials; material irradiation for advanced research in various scientific disciplines; and the applications with neutron beams. Through these objectives, RMB aims to enhance Brazil's technological development, and international competitiveness in the nuclear sector. Its radioisotope production includes a wide range of products, with emphasis on Mo-99 — which accounts for 85% of national demand—as well as Iridium-192 (Ir-192), Iodine-131 (I-131), Iodine-125 (I-125), Lutetium-177 (Lu-177), among others [10].

The RMB core will be configured in a 5 × 5 arrangement and will comprise 23 fuel elements made of $U_3Si_2$ dispersion-type in Aluminum, with a Uranium density of up to 3.5 gU/cm³ and an enrichment of 19.75% by weight of U-235 [10, 11], and two positions available for materials irradiation devices. This compact and efficient arrangement is designed to ensure high neutron flux at the core of $4 \times 10^{14}$ n/cm$^2$s, optimizing both radioisotope production and material irradiation experiments. The core is cooled and moderated by light water, with a downward forced flow and redundant cooling systems, ensuring thermal safety even under intensive operational conditions. The RMB design includes a Cold Neutron Source (CNS) consisting of 18 liters of

Liquid Ortho Deuterium at 23 K, located approximately 50 cm from the reactor core. The reactor's CNS will be responsible for moderating thermal neutrons, producing high-brightness beams and accessing neutrons with energies on the order of 3 meV. This will enable the implementation of techniques such as cold neutron imaging, reflectometry, and small-angle scattering [12]. The RMB's core configuration provides operational flexibility, allowing adjustments in irradiation setups and experiments for multiple purposes, such as nuclear fuel testing, material qualification, and the production of radioisotopes for medical and industrial applications. A schematic representation of the reflector tank and its coupled service pool is presented in Figure 2.

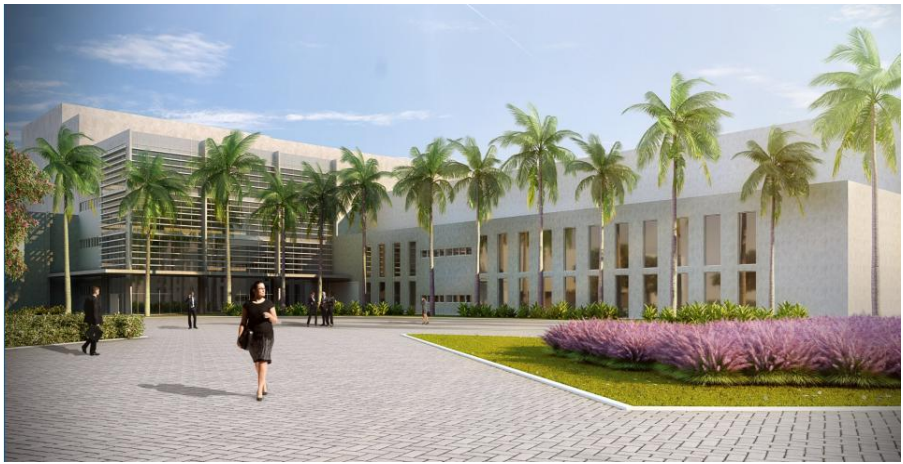

**Figure 1 – Artistic view of the entrance to the reactor building and neutron beam laboratory.**

The unit responsible for operating the neutron beamlines and instruments will be the National Neutron Beam Laboratory, equipped with an initial suite of 15 instruments [12]. These include high-resolution [13] and high-intensity powder diffractometers, strain-stress and single-crystal diffractometers, reflectometers, termal [14] and cold imaging instruments, SANS instruments [12], triple-axis spectrometers, and prompt-gamma instrument.

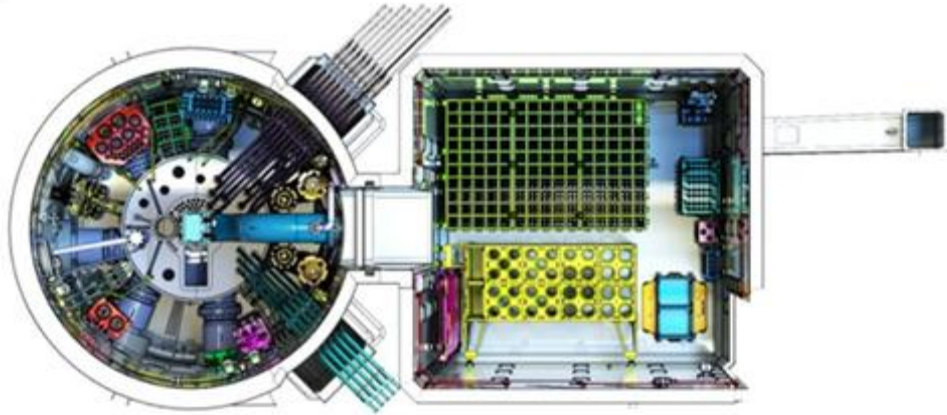

**Figure 2 – Detail of reactor (left) and service (right) pools.**

On the same site that houses the nuclear reactor, the project also includes administrative facilities, including restaurants, library an education center, offices for researchers and a hotel for visitors and users of the neutron beam laboratory. Alongside the RMB, the construction of a National Nuclear Fusion Laboratory, a Laser Technology Center, and a Ciclotron with high energy is also planned. Other research lines may utilize the established infrastructure for scientific investigations, such as the development of neutrino detectors and reactor core safeguards activities [15].

The purpose of this work is to present details of RMB and to discuss the versatility of its multipurpose potential. The demand for high neutron flux in all applications of a multipurpose research reactor can reduce the efficiency of physical processes such as neutron scattering experiments and chemical processes such as the production of radioisotopes with sufficient activity (measured in Ci) for medical and industrial applications. Therefore, we developed a methodology capable of evaluating the multipurpose nature of a research reactor through Monte Carlo simulations. We simulated neutron transport within the RMB reflector vessel (Figure 3) using the MCNP6 code and analyzed the neutron flux at critical positions—defined here as irradiation points furthest from the core—used for the production of Mo-99.

To assess the reactor's sensitivity to parallel process demands, we filled all irradiation units with Ir-191 and TeO$_2$ targets at maximum load. In addition, we simulated neutron absorption by dispersing Gadolinium-157 (Gd-157, an element with a high neutron absorption cross-section, $\sigma_{abs} = 2.54 \times 10^5$ barn [16]) in 6061 Aluminium (Al) alloy plates located in the dummy position. By varying the Gd-157 concentration (α), we were able to evaluate the neutron flux at critical positions, thereby suggesting a balance limit for the reactor's multipurpose capabilities. Given that the target's volume is the sum of the Al and Gd-157 volumes, $V_{target} = V_{Al} + V_{Gd-157}$, the number of Gd-157 and Al 6061 atoms are

$$N_{Gd-157} = \frac{\alpha V_{target} d_{Gd-157}}{M_{Gd-157}}, \quad N_{Al} = \frac{\beta V_{target} d_{Al}}{M_{Al}}, \qquad (eq.1)$$

where $\alpha + \beta = 1$, $d_{Gd-157}$, $M_{Gd-157}$, $d_{Al}$, $M_{Al}$ are the densities and molar masses of Gd-157 and Al, respectively. The 6061 Aluminium alloy, used in fuel assembly cladding, has a density of 2.71 g/cm³ and consists predominantly of Aluminium at 97.86% by weight. It contains 0.91% Magnesium, 0.64% Silicon, 0.25% Iron, and 0.21% Copper. Smaller amounts of other alloying elements include 0.05% Chromium, 0.03% Manganese, 0.03% Titanium, and 0.02% Zinc. The equivalent boron content is $1.21 \times 10^{-3}$. The values provided represent average concentrations, reflecting the typical composition of this alloy grade in the specified application.

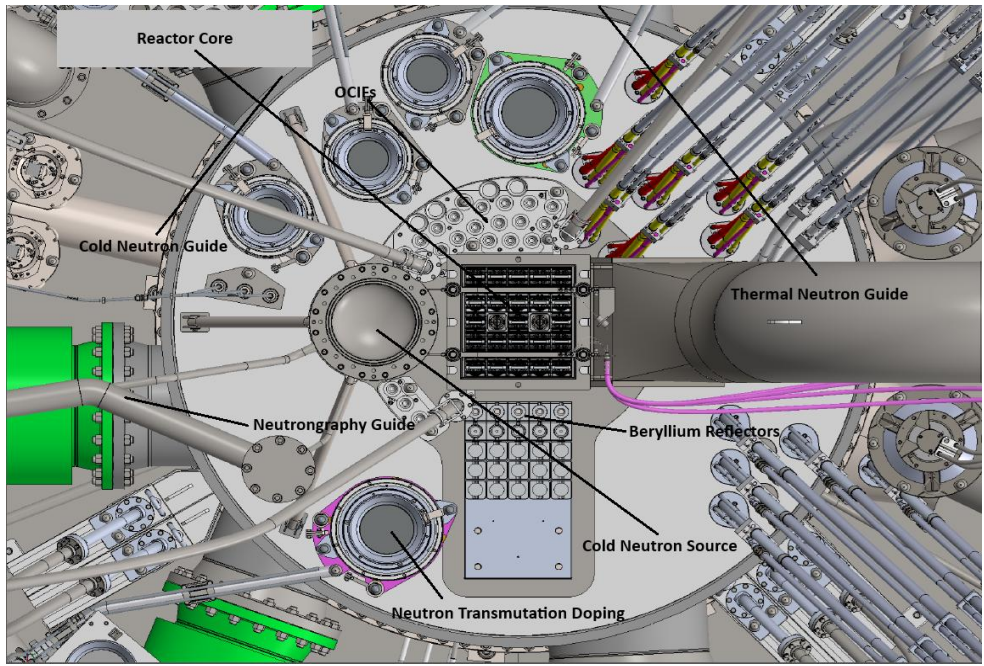

**Figure 3 – A view of RMB Reflector Vessel with some details.**

## 3 Some Remarks on the Graded Approach

The graded approach is a key principle in the nuclear field, enabling the adjustment of safety and design requirements according to the level of risk associated with each activity or component within a facility [1, 2]. In the case of the RMB, a multipurpose research reactor (supporting radioisotope production, neutron beam applications, and materials testing), this approach allows for the optimization of individual functions without compromising overall reactor safety. Components with higher radiological risk, such as irradiation targets, are subject to stricter controls, while lower-risk areas may be governed by proportionally reduced requirements.

The implementation of the graded approach in RMB is supported by simulation tools such as the MCNP6 code, which enable the assessment of how various operational demands affect neutron flux and reactor performance. This facilitates resource prioritization, enhances design flexibility, and promotes alignment with international standards. In summary, the graded approach provides a robust technical and regulatory framework to balance safety and efficiency in multipurpose research reactors like the RMB.

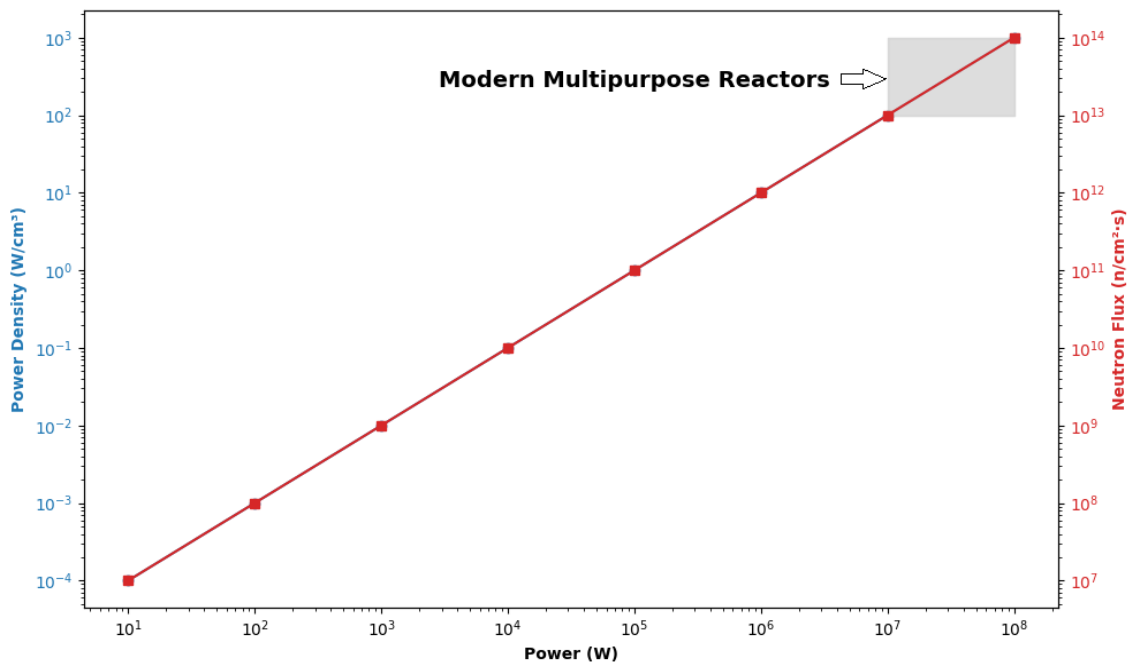

**Figure 4 - Relationship between power density, neutron flux, and reactor power [17]. It is observed that the gray rectangle corresponds to the requirements of modern multipurpose research reactors.**

In MNRs, the relationship between power, neutron flux, and power density is particularly significant, especially during transient conditions. In these reactors, the spatial distribution of neutron flux tends to be highly non-uniform, with regions of very high flux near the core and irradiation channels. During a transient event, such as a rapid power change or a reactivity pulse, this distribution dynamically changes, directly impacting the local power density — that is, the amount of energy released per unit volume at specific locations within the core. The relationship between flux and power, which is approximately linear under steady-state conditions (see Figure 4) [17], becomes more complex in this context, requiring careful spatial and temporal evaluation to avoid hot spots or damage to experimental targets.

RMB has a core designed to operate with a thermal power density of 312.5 W/cm$^3$ (assuming a core with an average volume of 40 cm × 40 cm × 60 cm), thereby providing a neutron flux in the range of $1.0 \times 10^{13}$ - $1.0 \times 10^{14}$ n/cm$^2$s. This parameter range, shown in gray in Figure 3, ensures that the RMB is classified as one of the world's modern MNRs.

**4 Monte Carlo Simulations of Reflector Vessel**

The reflector tank was modeled with the MCNP6 code [6], and simulations were carried out on the Coaraci Supercomputer, located at the University of Campinas (Unicamp) in Campinas, Brazil. Details on the construction of the reflector tank model can be found in reference [14].

RMB is equipped with three Out-of-Core Irradiation Facilities (OCIF) positions for producing Ir-192 for medical and industrial applications. The target's geometry—either seeds or pellets—is critical for its final application. We modeled both configurations to assess their impact. The seed model (S) consisted of 100 small cylinders (D = 1.9 mm, H = 2.0 mm), while the pellet model (P) used 72 larger discs (D = 2.7 mm, H = 18.0 mm). Both were simulated within a standard Aluminum holder (D = 31 mm, H = 94 mm). Our preliminary simulations revealed that the neutron flux is not globally affected by the choice of target geometry. This provides us the flexibility to use a simplified model that respects the maximum target load, which is appropriate for our study's focus on the reactor's maximum production limit. The TeO$_2$ targets, in turn, are modeled as cylinders with a 22 mm diameter and a 75 mm height, with each target being individually inserted into one of the 5 irradiation positions. The assemblies for Mo-99 production consist of a holder that secures multiple targets. Specifically, each holder contains four targets enclosed together within one capsule. Three such assemblies are loaded into each Mo-99 irradiation tube. The target's core ("meat") is a $UAl_2$-Al dispersion, while the cladding, holder, and capsule are all made of 6061 Aluminum alloy. Each target has a total Uranium mass of 8.962 g, which includes 1.77 g of Uranium-235 (U-235). The resulting Uranium density in the target core is approximately 2.388 g/cm³. A diagram showing the eleven irradiation positions for Mo-99 target production, the five positions for $TeO_2$ targets, the three positions for Ir-191 targets, and the dummy position containing Gd-157 is presented in Figure 5.

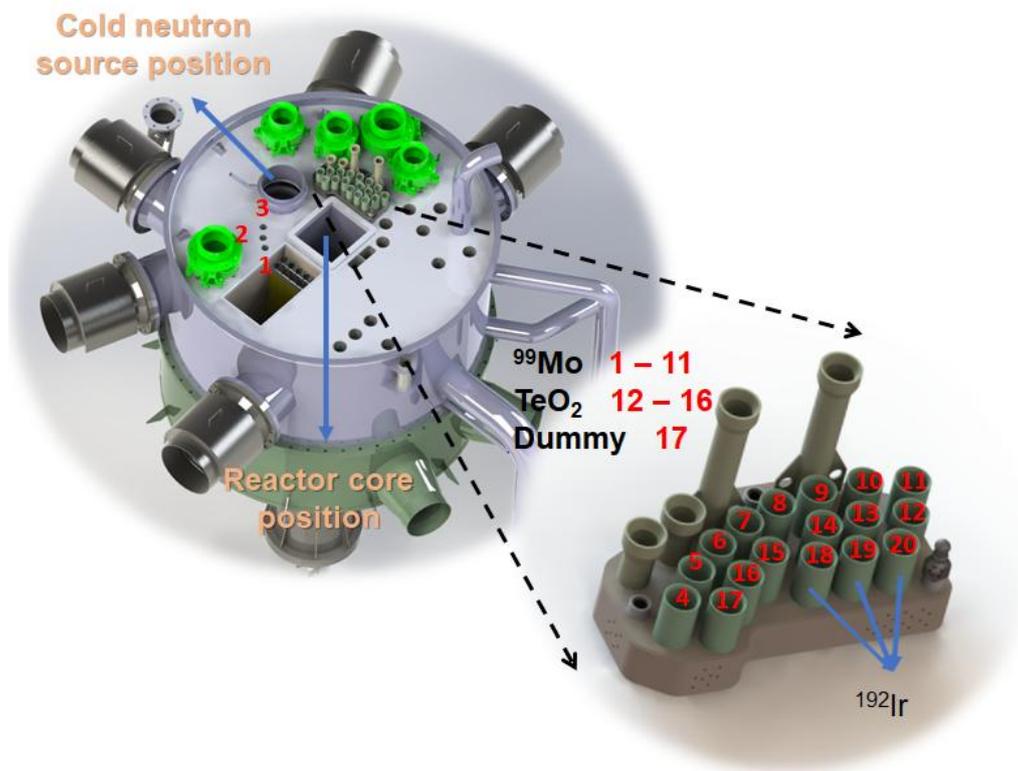

Figure 5 - Reflector tank and target positions: (#1–#11) Mo-99 production, (#12–#16) $TeO_2$, (#17) dummy containing Gd-157, and (#18–#20) for Ir-192 production.

On the other hand, the dummy OCIP, designed to simulate a void ("hole") in the reflector tank, shares the same structure as the Mo-99 target. The key difference between them is the dispersion of Gd-157 at various concentrations within the Aluminum plate, as detailed in Table 2.

Table 2 - Percentage of Gd-157 and delayed neutrons fraction in simulated cases.

| Case | %Gd-157 (α ∈ [0,1]) | β |
|------|---------------------|-----|
| 1    | 1                   | 0   |
| 2    | 0.9                 | 0.1 |
| 3    | 0.8                 | 0.2 |
| 4    | 0.7                 | 0.3 |
| 5    | 0.6                 | 0.4 |
| 6    | 0.5                 | 0.5 |
| 7    | 0.4                 | 0.6 |
| 8    | 0.3                 | 0.7 |
| 9    | 0.2                 | 0.8 |
| 10   | 0.1                 | 0.9 |
| 11   | 0.0                 | 1.0 |

It should be emphasized that the analysis developed in this study can be extended to cases in which the Gd-157 containing dummy is relocated within the reflector tank. A physical perturbation is expected around irradiation positions closest to the Gd-157 containing dummy, and this analysis will be carried out in future work. However, given the finalized design of RMB, we will focus on analyzing the effect of Gd-157 in a fixed position, dedicated to the irradiation of new targets.

At this stage, it is important to highlight that two distinct phases are analyzed in this work: the Beginning of Cycle (BOC) and the End of Cycle (EOC). In the BOC state, the reactor operates at an initial power of 30 MW, while the EOC state takes into account the fuel burnup after 7 days of operation, which is the time required for Mo-99 production. In practical terms, to maintain criticality, subsequent studies indicate the withdrawal of control rods by 85% to simulate the EOC in MCNP6 simulations. The boundary conditions that define this state are the solutions of the Bateman equation [18], considering thermal power limits of 80 kW in the Mo-99 targets. Therefore, the results obtained in this study are average values calculated for both BOC and EOC.

The activity of Mo-99 generated in a target depends on several parameters, including the irradiation time, the thermal neutron fission cross section of U-235, the thermal neutron flux incident on the material, the available mass of U-235, and the half-life of the radionuclide. Considering that the commercial activity of Mo-99 must be achieved within a maximum irradiation period of 7 days [19], it is possible to estimate the minimum required thermal neutron flux on the target. This estimation assumes a production efficiency consistent with that of major Mo-99-producing reactors, where the saturation time approaches the point of equilibrium between the Mo-99 production rate and its radioactive decay.

The accumulation of Mo-99 activity over time can be described by

$$A(t) = N\phi\sigma_f(1 - e^{-\lambda t}), \quad (eq.\ 2)$$

where $A(t)$ is the activity of Mo-99 in becquerels (Bq), $N$ is the number of U-235 atoms in the target, $\phi$ is the thermal neutron flux (n/cm²·s), $\sigma_f$ is the thermal fission cross section of U-235 (approximately $5.85 \times 10^{-24}$ cm²), and $\lambda$ is the decay constant of Mo-99, given by $\lambda = \frac{\ln(2)}{T_{1/2}}$, where $T_{1/2}$, which represents here the half-life of U-235, is given $T_{1/2} \approx 66$ hours [18]. Assuming an irradiation time of 7 days and considering a target containing 1.77 g of U-235, the number of atoms is approximately $N = 4.54 \times 10^{21}$, and targeting a commercial Mo-99 activity of 3000 $Ci$ (equivalent to $A(t = 7\ days) = 1.11 \times 10^{14}$ Bq), we can solve for the minimum neutron flux as $\phi_{min} = 5.1 \times 10^{12}$ n/cm²s. This value represents a conservative estimate based on typical operational conditions. Even under ideal irradiation conditions, only about 6% of the U-235 in the target is consumed during the irradiation cycle [19]. The remaining U-235, along with fission products and structural materials, must be managed as radioactive waste following the production process. For the purposes of this analysis, we will adopt the value of $\phi_{min}$ as the minimum thermal neutron flux required for Mo-99 production, taking into account the reactor's parallel operations, such as neutron beam extraction and the irradiation of other materials.

**5 Results and discussions**

The results of the neutron fluxes for all irradiation positions are presented in Tables 3 and 4, for the cases of beta equal to 1.0 and 0.0, respectively. It can be observed that the irradiation positions behave like points in an interconnected network, where the arrangement of targets seeks a balance between increasing and decreasing the neutron flux in first neighbors.

Even in the presence of a target composed entirely of Gd-157, the thermal neutron flux in the targets for Mo-99 production remains above the minimum calculated by equation (2), i.e., $\phi_{min} = 5.1 \times 10^{12}$ n/cm²s.

Table 3 - Results obtained for the case β = 1, that is, the dummy in position 17 is made only of Aluminum (%Gd-157 = 0). Maximum relative error = 3%.

| OCIF | Thermal Flux (n/cm²s) | Epithermal Flux (n/cm²s) | Fast Flux (n/cm²s) | Total Flux (n/cm²s) |
|---|---|---|---|---|
| Mo-#1 | 1.20734E+14 | 3.70124E+13 | 2.64223E+13 | 1.84168E+14 |
| Mo-#2 | 1.16024E+14 | 3.84878E+13 | 2.62800E+13 | 1.80792E+14 |
| Mo-#3 | 1.08016E+14 | 3.13382E+13 | 2.35073E+13 | 1.62862E+14 |
| Mo-#4 | 1.08421E+14 | 3.12181E+13 | 2.31132E+13 | 1.62752E+14 |
| Mo-#5 | 1.09583E+14 | 3.81999E+13 | 2.40411E+13 | 1.71824E+14 |
| Mo-#6 | 9.75348E+13 | 3.67532E+13 | 2.24464E+13 | 1.56734E+14 |
| Mo-#7 | 8.52491E+13 | 3.09946E+13 | 1.96175E+13 | 1.35861E+14 |
| Mo-#8 | 7.85337E+13 | 2.62302E+13 | 1.76633E+13 | 1.22427E+14 |
| Mo-#9 | 7.38239E+13 | 2.52416E+13 | 1.69315E+13 | 1.15997E+14 |
| Mo-#10 | 7.79808E+13 | 2.58087E+13 | 1.78876E+13 | 1.21677E+14 |
| Mo-#11 | 9.21167E+13 | 2.85964E+13 | 2.05439E+13 | 1.41257E+14 |
| TeO$_2$-#12 | 1.38526E+14 | 3.10240E+13 | 4.38596E+12 | 1.73936E+14 |
| TeO$_2$-#13 | 1.21725E+14 | 2.54617E+13 | 4.31922E+12 | 1.51506E+14 |
| TeO$_2$-#14 | 1.20446E+14 | 2.63297E+13 | 4.52606E+12 | 1.51302E+14 |
| TeO$_2$-#15 | 1.27384E+14 | 3.09566E+13 | 4.87621E+12 | 1.63217E+14 |
| TeO$_2$-#16 | 1.56115E+14 | 4.66314E+13 | 6.38666E+12 | 2.09133E+14 |
| dummy-#17 | 2.08290E+14 | 5.37563E+13 | 5.99008E+12 | 2.68037E+14 |
| Ir-#18 | 3.83463E+13 | 6.40764E+13 | 1.14767E+13 | 1.13899E+14 |
| Ir-#19 | 3.99135E+13 | 6.72665E+13 | 1.19737E+13 | 1.19154E+14 |
| Ir-#20 | 4.05584E+13 | 6.58226E+13 | 1.13484E+13 | 1.17729E+14 |

Table 4 - Results obtained for the case β = 0, that is, the dummy in position #17 is made only of Gd-157. Maximum relative error = 3%.

| OCIF | Thermal Flux (n/cm²s) | Epithermal Flux (n/cm²s) | Fast Flux (n/cm²s) | Total Flux (n/cm²s) |
|---|---|---|---|---|
| Mo-#1 | 1.20515E+14 | 3.70690E+13 | 2.64360E+13 | 1.84020E+14 |
| Mo-#2 | 1.15689E+14 | 3.88288E+13 | 2.62548E+13 | 1.80772E+14 |
| Mo-#3 | 1.06310E+14 | 3.12773E+13 | 2.33744E+13 | 1.60962E+14 |
| Mo-#4 | 9.50240E+13 | 2.73234E+13 | 2.02392E+13 | 1.42587E+14 |
| Mo-#5 | 8.36025E+13 | 3.08742E+13 | 1.86641E+13 | 1.33141E+14 |
| Mo-#6 | 7.07301E+13 | 2.94037E+13 | 1.65199E+13 | 1.16654E+14 |
| Mo-#7 | 6.92186E+13 | 2.60938E+13 | 1.60642E+13 | 1.11377E+14 |
| Mo-#8 | 7.05071E+13 | 2.38608E+13 | 1.59762E+13 | 1.10344E+14 |
| Mo-#9 | 6.95603E+13 | 2.36630E+13 | 1.59688E+13 | 1.09192E+14 |
| Mo-#10 | 7.58388E+13 | 2.51486E+13 | 1.73179E+13 | 1.18305E+14 |
| Mo-#11 | 9.12553E+13 | 2.81215E+13 | 2.01011E+13 | 1.39478E+14 |
| TeO$_2$-#12 | 1.41433E+14 | 3.16592E+13 | 4.43574E+12 | 1.77528E+14 |
| TeO$_2$-#13 | 1.22413E+14 | 2.55805E+13 | 4.31554E+12 | 1.52309E+14 |
| TeO$_2$-#14 | 1.18298E+14 | 2.60919E+13 | 4.46151E+12 | 1.48851E+14 |
| TeO$_2$-#15 | 1.18032E+14 | 3.00780E+13 | 4.70440E+12 | 1.52814E+14 |
| TeO$_2$-#16 | 1.27381E+14 | 4.40647E+13 | 6.03011E+12 | 1.77475E+14 |
| dummy-#17 | 2.06735E+13 | 4.43907E+13 | 5.20609E+12 | 7.02703E+13 |
| Ir-#18 | 3.70388E+13 | 6.33573E+13 | 1.09418E+13 | 1.11338E+14 |
| Ir-#19 | 3.96952E+13 | 6.58440E+13 | 1.13044E+13 | 1.16844E+14 |
| Ir-#20 | 4.09495E+13 | 6.71979E+13 | 1.15037E+13 | 1.19651E+14 |

It can be observed in Figures 6 and 7 that there is only a slight alteration on the neutron flux map due the presence of a highly absorbing material. It can also be concluded that the dimensions of the irradiation targets guarantee the stability of the reactor's operation, ensuring the production of radioisotopes, irradiation of materials, and extraction of neutron beams.

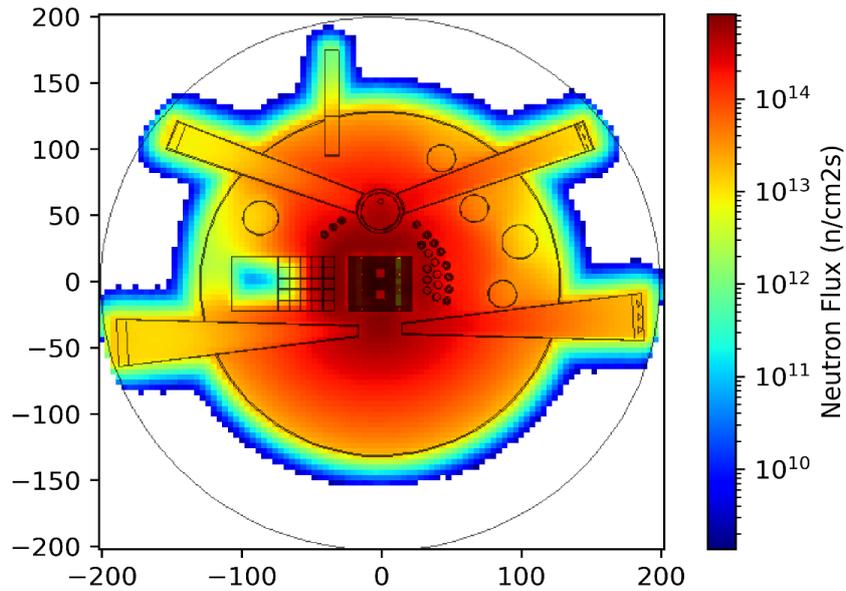

**Figure 6 - Neutron flux map generated by MCNP6, obtained for the case β = 1, that is, the dummy in position #17 is made only of Al (%Gd-157 = 0).**

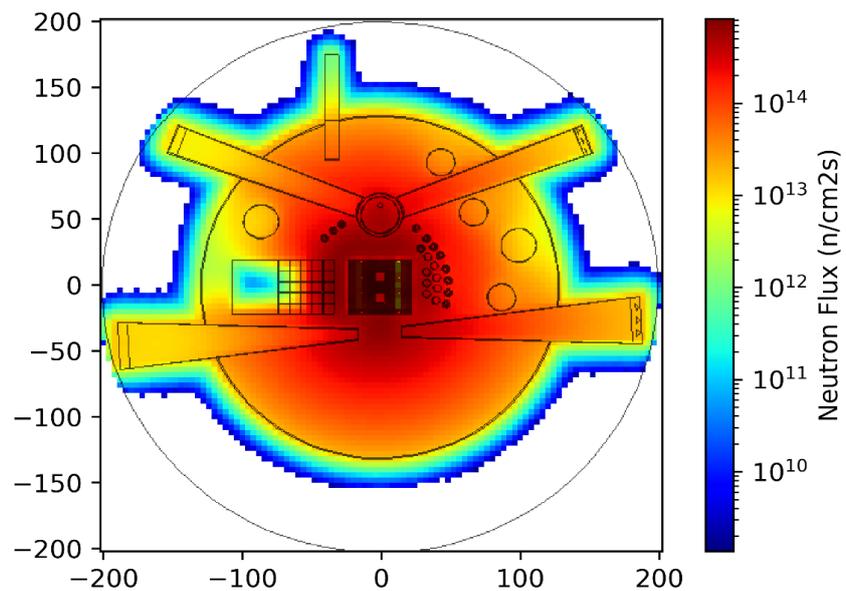

**Figure 7 - Neutron flux map generated by MCNP6, obtained for the case β = 0, that is, the dummy in position #17 is made only of Gd-157 (%Al = 0).**

The definitions in equations (3-4) provide us with parameters to measure the influence of the Gd-157 target on other fixed irradiation capsules (OCIFs), where the values in red are highlighted. It should be noted that the U-235 targets in positions #4-9 exhibit the largest deviations (Table 5). Although a local perturbation exists, there are no global effects that would prevent the objective of irradiating all targets in the RMB.

$$\eta_{thermal} = \frac{\phi_{thermal}(\beta = 0) - \phi_{thermal}(\beta = 1)}{\phi_{thermal}(\beta = 1)} \times 100, \qquad (eq.3)$$

$$\eta_{epithermal} = \frac{\phi_{epithermal}(\beta = 0) - \phi_{epithermal}(\beta = 1)}{\phi_{epithermal}(\beta = 1)} \times 100. \qquad (eq.4)$$

Table 5 - Results for $\eta_{thermal}$ and $\eta_{epithermal}$, with values highlighted in red indicating variations exceeding a statistical error of 3%.

| OCIF | $\eta_{thermal}$ (%) | $\eta_{epithermal}$ (%) |
|---|---|---|
| Mo-#1 | -0.18 | 0.15 |
| Mo-#2 | -0.29 | 0.89 |
| Mo-#3 | -2.39 | -0.19 |
| Mo-#4 | **-12.36** | **-12.48** |
| Mo-#5 | **-23.71** | **-19.18** |
| Mo-#6 | **-27.48** | **-20.00** |
| Mo-#7 | **-18.80** | **-15.81** |
| Mo-#8 | **-10.22** | **-9.03** |
| Mo-#9 | **-5.78** | **-6.25** |
| Mo-#10 | -2.75 | -2.56 |
| Mo-#11 | -0.94 | -1.66 |
| TeO$_2$-#12 | 2.10 | 2.05 |
| TeO$_2$-#13 | 0.57 | 0.47 |
| TeO$_2$-#14 | -1.78 | -0.90 |
| TeO$_2$-#15 | **-7.34** | 2.84 |
| TeO$_2$-#16 | **-18.41** | **-5.50** |
| dummy-#17 | **-90.10** | **-17.42** |
| Ir-#18 | **-3.41** | -1.12 |
| Ir-#19 | -0.55 | -2.11 |
| Ir-#20 | 0.66 | 2.09 |

Individual results for the three different irradiated materials can be seen in Figures 8-11, representing the ratio between the thermal neutron fluxes at the irradiation positions in the absence and presence of Gd-157 in the dummy, respectively, in function of Al concentration. This ratio allows the effect of Gd-157 to be interpreted as irradiation efficiency, given that the MCNP6 code provides the neutron flux that passes through the volume delimited by the target. Therefore, higher values for the neutron flux indicate low absorption values in the volume in question. Consequently, ratios greater than 1.0 suggest that the arrangement creates a compensating movement, between absorbing and reflecting neutrons at positions closer to and further from the reactor core. The termal neutron capture cross-sections for U-235, Ir-191, and Te-130 are $\sigma_{U-235} = 98.8\ barn$, $\sigma_{Ir-191} = 954\ barn$, and $\sigma_{Te-130} = 0.27\ barn$, respectively [20]. Consequently, due to the neutron absorption effect from the adjacent Gd-157 target (position #17), the Mo-99 production targets at positions #4-#7 exhibit the greatest perturbations (10 - 40%) for 0 < β < 1.0. Due to its low absorption cross-section, Te-130 results in a smaller perturbation in the irradiation positions compared to U-235 (Figure 9), except for position #16 (18.41%) adjacent to the dummy. Meanwhile, Figure 10 displays minor perturbations at positions #18-#20 for 0.0 < β < 1.0.

The fit in Figure 11 allows us to obtain a relation for the thermal neutron flux ($\phi_{dummy}(\beta)$) passing through the dummy as a function of its composition, that is, the percentage of Al and Gd-157, with $R^2 = 0.9999$,

$$\phi_{dummy}(\beta) = \frac{7.54674 \times 10^{13}}{4.65217 e^{-1.24\beta} - 1}. \qquad (eq.5)$$

It is interesting to note that the resulting expression (eq. 5) is the typical form for the problem of flux through a body as a function of its thickness [18], which in our case is emulated by the β parameter.

Although Gd-157 has a high cross section, no irradiation point undergoes sufficient changes to invalidate the final objective of the reactor. The maximum perturbation found (27.48%) at position #6 does not drastically affect the production of Mo-99, given that $\phi_{min} = 5.1 \times 10^{12}$ n/cm²s. This is an interesting result, as it allows the RMB to propose irradiating elements with high neutron absorption without compromising parallel activities of the reactor. That is, RMB stands out in the radioisotope production scenario due to the great potential for applications of MNRs.

The results also have implications for the production of Terbium (Tb) isotopes in MNRs. Isotopes such as Tb-155 and Tb-156 can be produced by the irradiation of Gd-157 targets with protons or deuterons in a nuclear reactor, through reactions such as Gd-157(p,3n)Tb-155 and Gd-157(d,4n)Tb-155, respectively. Although these are not traditional production methods for nuclear reactors, the production of Tb-160 from natural Gadolinium via the neutron activation of Gd-158 is well-established [21].

Due to their applications in nuclear medicine, Terbium isotopes have garnered considerable attention from the international community [22]. In this study, Gd-157 targets were chosen because it is the isotope with the largest neutron absorption cross-section, effectively serving as a test of the reactor's stability to ensure its integrity during parallel operations. Nevertheless, the irradiation of other Gd isotopes is a readily implementable pathway that unlocks significant potential for applications in MNRs.

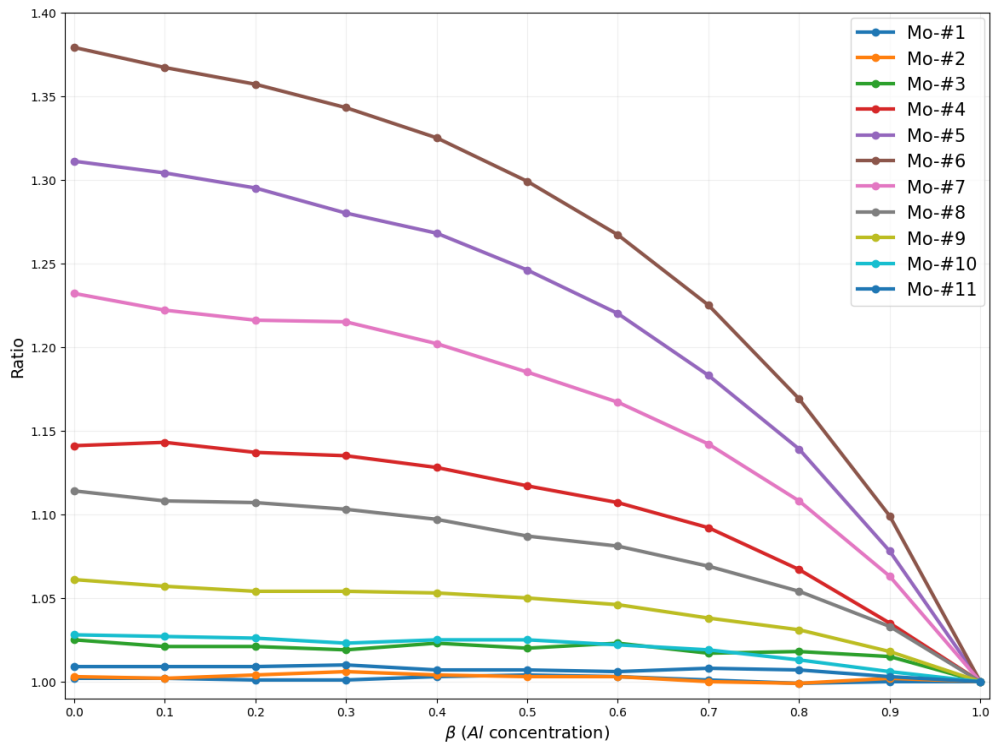

**Figure 8 – Results for the target positions of Mo-99 production. The vertical axis represents the ratio between the thermal neutron fluxes reaching the target positions for the case β = 1 and an arbitrary β, respectively.**

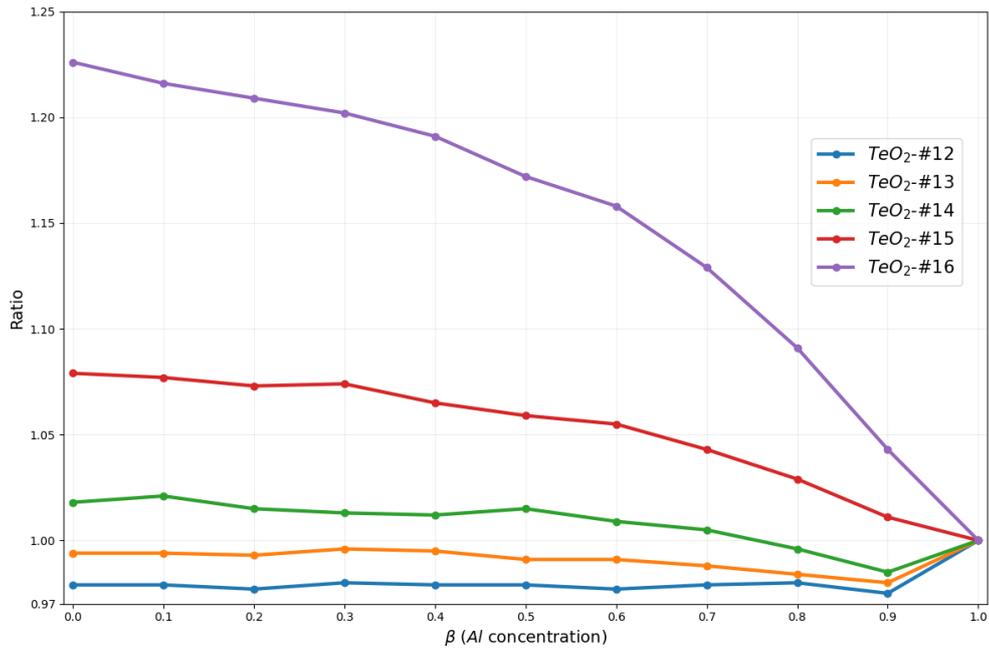

**Figure 9 – Results for the target positions for I-131 production. The vertical axis represents the ratio between the thermal neutron fluxes reaching the target positions for the case β = 1 and an arbitrary β, respectively.**

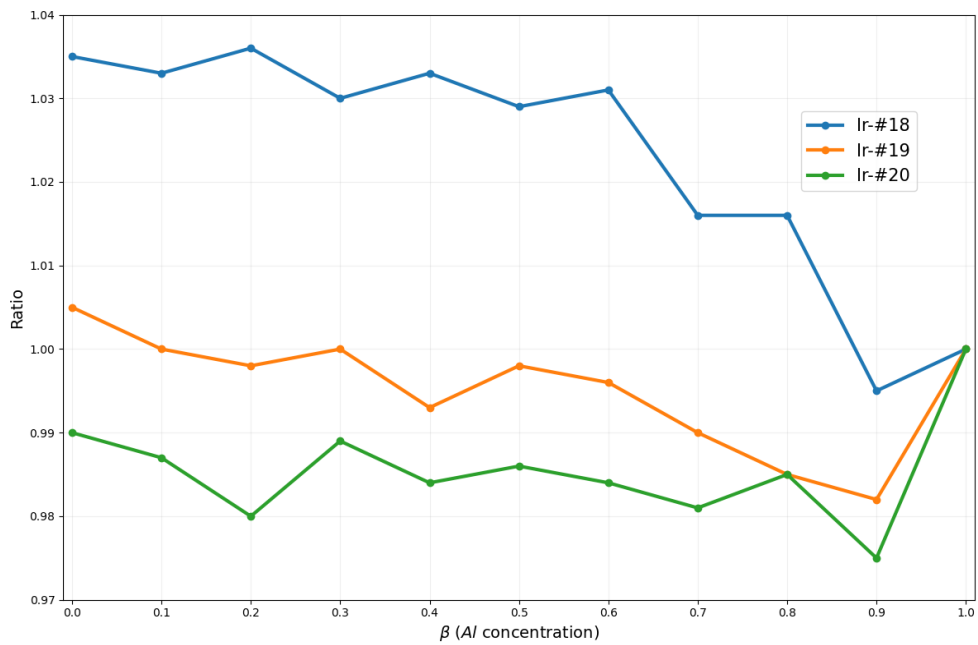

**Figure 10 – Results for the target positions of Ir-192 production. The vertical axis represents the ratio between the thermal neutron fluxes reaching the target positions for the case β = 1 and an arbitrary β, respectively.**

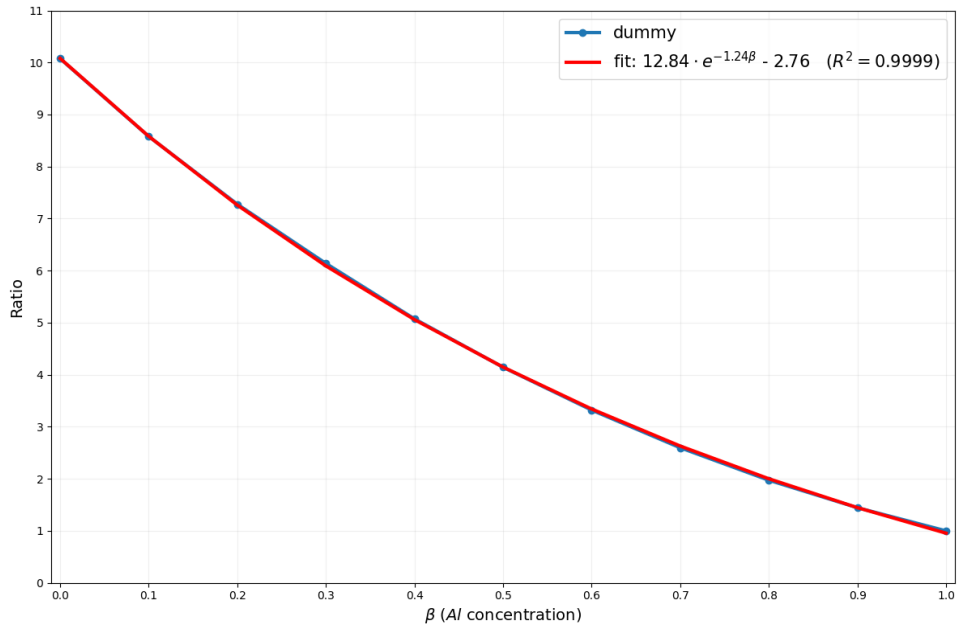

**Figure 11 – Results of dummy position. The vertical axis represents the ratio between the thermal neutron fluxes reaching for the case β = 1 and an arbitrary β, respectively.**

Gracheva and colleagues demonstrated that the isotope Tb-161 is a promising radionuclide for cancer therapy, exhibiting decay properties similar to the clinically applied Lu-177 but with enhanced therapeutic potential due to the additional emission of conversion and Auger electrons [23]. They also reported the production of Tb-161 via irradiation of Gd-160 targets followed by purification, yielding Tb-161Cl$_3$ with ≥ 99% purity and suitable for stable radiolabeling of DOTA, a chelator that binds the radionuclide, and DOTATOC, a target peptide linked to DOTA that directs the radionuclide to tumors expressing somatostatin receptors, highlighting its potential for future clinical applications.

In an experimental context, Muller and colleagues evaluated the long-term renal effects of Tb-161-folate administration, comparing them with those induced by Lu-177-folate [24]. Tb-161-folate caused dose-dependent radionecropathy over time but did not result in more severe renal damage than Lu-177-folate administered at the same activity. These findings suggest that the additional conversion and Auger electrons do not exacerbate overall renal injury compared to Lu-177-folate, although they do contribute to increased dose deposition in renal tissue. The authors emphasized that the potential systemic toxicity affecting tissues beyond the kidneys still needs to be investigated following Tb-161-based therapy.

Given the potential applications in nuclear medicine, our initial results indicate that the irradiation of Gd-160 ($\sigma_{abs} = 1.4$ barn [16])) targets is feasible within the Brazilian MNR facilities, based on the reactor's stability when exposed to a highly neutron-absorbing material such as Gd-157. Neutron capture by Gd-160 produces the isotope Gd-161, which, through beta decay, provides a production route for Tb-161 [24]. Therefore, we can conclude that the RMB has significant potential to produce a wide variety of radioisotopes without compromising the reactor's other applications, demonstrating its multipurpose character.

## 6 Summary and conclusions

This study evaluates the impact of irradiating a highly neutron-absorbing material, Gd-157, on the operational stability and parallel activities of a multipurpose research reactor. The results demonstrate that the reactor's neutron flux remains remarkably stable, with its irradiation targets behaving as an interconnected network that effectively balances local perturbations. Even with a pure Gd-157 target, the flux required for key processes, such as Mo-99 production, is maintained above the necessary minimum. This confirms the reactor's design has robustness operational, demonstrating that it can reliably perform parallel tasks without being compromised by the simultaneous irradiation of highly absorbing materials. The successful test validates the reactor's robust design and unlocks new possibilities for research, including the irradiation of various Gadolinium isotopes to create valuable Terbium radioisotopes for nuclear medicine. Ultimately, these results underscore the reactor's significant potential and flexibility, positioning it as a key asset in the landscape of radioisotope production.

## 7 Acknowledgments

We thank the Coaraci Supercomputer for computer time (Fapesp grant #2019/17874-0) and the Center for Computing in Engineering and Sciences at Unicamp (Fapesp grant #2013/08293-7). We would like to thank M.Sc. Antonio C. Igleisias for running our Monte Carlo simulations. C.G.S. Santos and I.S.R. Júnior thank CNEN/Fundação PATRIA for the research scholarships 006/2025 and 007/2025, respectively, under the FINEP agreement 01.24.0373.00 (RMB280).

**Conflict of Interest**

There is no conflict of interest.

**References**

[1] INTERNATIONAL ATOMIC ENERGY AGENCY. *Safety of Research Reactors*. IAEA Safety Standards Series No. NS-R-4, IAEA, Vienna (2005).

[2] INTERNATIONAL ATOMIC ENERGY AGENCY. *Use of a graded approach in the application of the safety requirements for research reactors*. IAEA TECDOC Series No. 1672, Vienna (2012).

[3] SEO, C. G.; CHO, N. Z. A core design concept for multipurpose research reactors. *Nuclear Engineering and Design*, **252**, 34–41 (2012). https://doi.org/10.1016/j.nucengdes.2012.06.031.

[4] RAINA, V. K.; SASIDHARAN, K.; SENGUPTA, S.; SINGH, T. Multipurpose research reactor. *Nuclear Engineering and Design*, **236**(7–8), 770–783 (2006). https://doi.org/10.1016/j.nucengdes.2005.09.022.

[5] TERUEL, F. E.; UDDIN, R. An innovative research reactor design. *Nuclear Engineering and Design*, **239**(2), 395–407 (2009). https://doi.org/10.1016/j.nucengdes.2008.10.025.

[6] GOORLEY, T. et al. Features of MCNP6. *Annals of Nuclear Energy*, **87**, 772 (2016).

[7] VILLARINO, E.; DOVAL, A. INVAP's research reactor designs. *Science and Technology of Nuclear Installations*, **2011**, 1–12 (2011). https://doi.org/10.1155/2011/490391.

[8] NGUYEN, N. D.; NGUYEN, K. C.; HUYNH, T. N.; VO, D. H. D.; TRAN, H. N. Conceptual design of a 10 MW multipurpose research reactor using VVR-KN fuel. *Science and Technology of Nuclear Installations*, **2020**, Article ID 7972827, 11 pages. https://doi.org/10.1155/2020/7972827

**[9]** TARTAGLIONE, A. et al. Present and future activities on neutron imaging in Argentina. *Physics Procedia*, **69**, 142–149 (2015). https://doi.org/10.1016/j.phpro.2015.07.018.

**[10]** PERROTTA, J. A.; SOARES, A. J. RMB: The new Brazilian multipurpose research reactor. *ATW. International Journal for Nuclear Power*, **60**, 30–34 (2015).

**[11]** CAMPOLINA, D.; DA COSTA, A. C. L.; ANDRADE, E. P.; SANTOS, A. A. C.; VASCONCELOS, V. Neutronic analysis of the fuel loaded irradiation loop device of the RMB Multipurpose Brazilian Reactor. *Progress in Nuclear Energy*, **104**, 109–116 (2018). https://doi.org/10.1016/j.pnucene.2017.09.006.

**[12]** SOUZA, A. P. S.; DE OLIVEIRA, L. P.; YOKAICHIYA, F.; GENEZINI, F. A.; FRANCO, M. K. K. D. Neutron guide building instruments of the Brazilian Multipurpose Reactor (RMB) project. *Journal of Instrumentation*, **15**, P04011 (2020). https://doi.org/10.1088/1748-0221/15/04/P04011.

**[13]** SOUZA, A. P. S.; DE OLIVEIRA, L. P.; GENEZINI, F. A.; SANTOS, A. Simulating Araponga – the high-resolution diffractometer of Brazilian Multipurpose Reactor. *Brazilian Journal of Radiation Sciences*, **10**(3B Suppl.), (2022). https://doi.org/10.15392/2319-0612.2022.1861.

**[14]** DE OLIVEIRA, L. P.; SOUZA, A. P. S.; GENEZINI, F. A.; DOS SANTOS, A. Stochastic modeling of a neutron imaging center at the Brazilian Multipurpose Reactor. *Nuclear Engineering and Design*, **420**, 113042 (2024). https://doi.org/10.1016/j.nucengdes.2024.113042.

**[15]** DE OLIVEIRA, L. P. Estimating the detection of antineutrinos in the future Brazilian neutron source. *arXiv preprint*, arXiv:2403.17812v3 (2024). https://arxiv.org/abs/2403.17812.

**[16]** DUMAZERT, J. *et al.* Gadolinium for neutron detection in current nuclear instrumentation research: A review. Nuclear Instruments and Methods in Physics Research Section A: Accelerators, Spectrometers, Detectors and Associated Equipment **882**, pp 53-68, (2018). https://doi.org/10.1016/j.nima.2017.11.032

**[17]** PERROTTA, J. A. *Considerations and Challenges of Research Reactor Management.* In: INTERNATIONAL ATOMIC ENERGY AGENCY. *Research Reactors: Addressing Challenges and Opportunities to Ensure Effectiveness and Sustainability: Proceedings of an International Conference,* Buenos Aires, Argentina, 25–29 Nov. 2019. Vienna: IAEA, 2022. (IAEA-CN–277/O.7.06). Available at: https://www.iaea.org/publications/14761/research-reactors-addressing-challenges-and-opportunities-to-ensure-effectiveness-and-sustainability

**[18]** LEWIS, E. E. *Fundamentals of Nuclear Reactor Physics*. 1st ed. Amsterdam: Academic Press, 312 p. (2008). ISBN 978-0-12-370631-7.

**[19]** NATIONAL RESEARCH COUNCIL (U.S.). *Medical isotope production without highly enriched uranium*. Washington, DC: National Academies Press, 2009. Available at: https://doi.org/10.17226/12569.

**[20]** BROWN, D. A. *et al.* ENDF/B-VIII.0: The 8th Major Release of the Nuclear Reaction Data Library with CIELO-project Cross Sections, New Standards and Thermal Scattering Data. Nuclear Data Sheets **148**, p. 1-142, (2018). https://doi.org/10.1016/j.nds.2018.02.001

**[21]** WANG, Y. *et al.* Study of terbium production from enriched Gd targets via the reaction 155Gd(d,2n)155Tb. Applied Radiation and Isotopes, **201**, p. 110996, (2023). https://doi.org/10.1016/j.apradiso.2023.110996

**[22]** PYLES, J. M. *et al.* Production and purification of Terbium-155 using natural gadolinium targets. ACS Omega, **10**, n. 28, p. 30335-30343, (2025). https://doi.org/10.1021/acsomega.5c01653

**[23]** GRACHEVA, N. *et al.* Production and characterization of no-carrier-added $^{161}$Tb as an alternative to the clinically-applied $^{177}$Lu for radionuclide therapy. *EJNMMI radiopharm. chem.* **4**, 12 (2019). https://doi.org/10.1186/s41181-019-0063-6

**[24]** MÜLLER, C. *et al.* $^{161}$Tb-folate: a potential theranostic agent for folate receptor-positive cancer. Nuclear Medicine and Biology **43**, n. 1-2, p. 1-8, (2016). https://doi.org/10.1016/j.nucmedbio.2015.10.007